\documentclass[twocolumn,showpacs]{revtex4}

\usepackage{graphicx}
\usepackage{dcolumn}
\usepackage{amsmath}
\usepackage{color}

\begin{document}

\title{
Chiral sine-Gordon model
}

\author{Takashi Yanagisawa}

\affiliation{Electronics and Photonics Research Institute,
National Institute of Advanced Industrial Science and Technology (AIST),
Central 2, 1-1-1 Umezono, Tsukuba, Ibaraki 305-8568, Japan
}

\date{February 15 2016}

\begin{abstract}
We investigate the chiral sine-Gordon model using the renormalization
group method.
The chiral sine-Gordon model is a model for $G$-valued fields and describes 
a new class of phase transitions, where $G$ is a compact Lie group.
We show that the model is renormalizable by means of a perturbation expansion
and we derive beta functions of the renormalization group theory.
The coefficients of beta functions are represented by the Casimir invariants.
The model contains both asymptotically free and ultraviolet strong coupling 
regions.
The beta functions have a zero which is a bifurcation point that divides 
the parameter
space into two regions; they are the weak coupling region and the strong
coupling region.
A large-$N$ model is also considered. This model is reduced to the 
conventional sine-Gordon model that
describes the Kosterlitz-Thouless transition near the fixed point.
In the strong-coupling limit, the model is reduced to a $U(N)$ matrix
model.
\end{abstract}


\maketitle

{\bf Introduction}
The chiral model was generalized to the Wess-Zumino-Witten (WZW) model
by including the Wess-Zumino term.\cite{wes71,wit83,wit84,nov82,nov82b}.
The WZW model gives a model of conformal field theory whose current
algebra is realized by a Kac-Moody algebra.
The massive chiral model, which is the chiral Lagrangian\cite{gol78,bre79}
or the non-linear sigma model with the mass term, is also interesting as a
two-dimensional field theory.
The massive chiral model can be regarded as a generalization of the
sine-Gordon model to the model for $G$-valued fields\cite{nit15},
where we add the term of type ${\rm Tr}(g+g^{-1})$ for $g\in G$ where 
$G$ is a general gauge group (Lie group).

The sine-Gordon model has universality and appears in various fields
of physics\cite{das75,zam79,raj82,man04}.
The two-dimensional (2D) sine-Gordon model describes the Kosterlitz-Thouless
transition of the 2D classical XY model\cite{kos73,kos74}.
The 2D sine-Gordon model is mapped to the Coulomb gas model with
logarithmic interaction\cite{zin89}.
The Kondo problem also belongs to the same universality class where the
scaling equations are given by those for the 2D sine-Gordon model\cite{kon12}.
The one-dimensional Hubbard model is also mapped onto the 2D sine-Gordon
model by using a bosonization method\cite{sol79,hal81}.
The sine-Gordon model plays an important role in superconductors,
especially multi-band 
superconductors\cite{leg66,tan10a,tan10b,yan12,yan13,yan14} 
including layered 
high-temperature superconductors.
Generalized sine-Gordon models have also been 
investigated\cite{par96,bak96,car97}.

In this paper we investigate the $G$-valued nonabelian sine-Gordon model.
We consider compact Lie groups such as $G=SU(N)$.
The nonabelian sine-Gordon model is renormalizable as for the $U(1)$ 
sine-Gordon model.
Although the chiral model shows an asymptotic freedom in two dimensions,
it is lost by the mass term in general.
The beta functions, however, have zero at a critical point and
this point is a bifurcation point that divides the parameter space
into two regions; one is the weak coupling region and the other is
the strong coupling region.
The asymptotic freedom is realized in the weak coupling region.
In the strong coupling limit, the $SU(N)$ sine-Gordon model is
reduced to a unitary matrix model.  It has been shown by Gross and
Witten that in the large $N$ limit there is a third-order transition
at some critical coupling constant\cite{gro80}.
Brezin and Gross generalized the coupling constant to be a matrix
and found that there is also a phase transition\cite{bre80, bro81, bre10}.

This paper is organized as follows.
In Section II, we show the action of the $G$-valued chiral 
sine-Gordon model.
In section III, we present the renormalization group method in the
minimal subtraction scheme using
the dimensional regularization method\cite{bol72,tho72,gro75}.
We discuss the renormalization flow in Section IV.
In the subsequent section we discuss a relation to the 
Kosterlitz-Thouless transition for large $N$.  We also examine a
relationship with a unitary Matrix model in the strong coupling limit.
We give a discussion on the relevance of our model to some problems and
give a summary in the last Section.

{\bf Chiral sine-Gordon model}
The sine-Gordon model is given by\cite{raj82,zin89}
\begin{equation}
\mathcal{L}= \frac{1}{2t}\left(\partial_{\mu}\varphi\right)^2
+\frac{\alpha}{t}\cos\varphi,
\end{equation}
where $\varphi$ is a real scalar field.
This Lagrangian is written as, by defining $g=e^{i\varphi}\in U(1)$, 
\begin{equation}
\mathcal{L}= \frac{1}{2t}\partial_{\mu}g\partial^{\mu}g^{-1}
+\frac{\alpha}{2t}(g+g^{-1}).
\end{equation}
This is generalized to a general Lie group $G$:
\begin{equation}
\mathcal{L}= \frac{1}{2t}{\rm Tr}\partial_{\mu}g\partial^{\mu}g^{-1}
+\frac{\alpha}{2t}{\rm Tr}(g+g^{-1}),
\end{equation}
for $g\in G$.  We adopt that $t>0$ and $\alpha>0$.
This model can be regarded as a chiral model with the potential term
of sine-Gordon type.
In this paper we consider the chiral sine-Gordon model and derive the 
renormalization group equation.
The renormalization group equation for the sine-Gordon model was
derived by using the Wilson method\cite{kog79} or the perturbation 
method\cite{zin89}. 
In this paper we use the perturbation in terms of $t$ and $\alpha$.

{\bf Renormalization Group Theory}
{\em Chiral Lagrangian}
An element $g$ of the Lie Group $G$  is represented in the form:
\begin{equation}
g= g_0\exp(i\lambda T_a\pi_a),
\end{equation}
where $\lambda$ is a real number $\lambda\in {\bf R}$ and $g_0$ is a
some element in $G$.
Repeated indices imply the summation is to be done.
$\{T_a\}$ ($a=1,2,\cdots, N_T$) is a basis set of the Lie algebra ${\bf g}$
where ${\bf g}$ is the Lie algebra of $G$.
$N_T=N^2-1$ for $G=SU(N)$.
$\pi_a$ ($a=1,2,\cdots,N_T$) are scalar fields.
$\lambda$ is introduced as an expansion parameter and the results
do not depend on $\lambda$.  Thus we can put $\lambda=1$.
$\{T_a\}$ satisfy
\begin{equation}
\left[ T_a,T_b\right]=if_{abc}T_c,
\end{equation}
where $\{f_{abc}\}$ are structure constants and are totally antisymmetric.
$\{T_a\}$ are normalized as
\begin{equation}
{\rm Tr}T_aT_b= c\delta_{ab},
\end{equation}
for a real constant $c$.
The normalization constant can take on any real number.
For example,
for $G=SU(2)$ we can take $T_a=\sigma_a/2$ (Pauli matrices)
with $c=1/2$. 

The renormalization of the coupling constant $t$ comes from
the kinetic term and the mass term.
The former is missing for the conventional ($U(1)$) sine-Gordon
model.
Let us first consider the renormalization of the kinetic term,
namely,  the chiral 
Lagrangian given as
\begin{equation}
\mathcal{L}_{chiral}=
\frac{1}{2t_0}{\rm Tr}\partial_{\mu}g\partial^{\mu}g^{-1},
\end{equation}
where $t_0$ is the bare coupling constant.
We define the renormalized coupling constant $t$ in the following way:
\begin{equation}
t_0= t\mu^{2-d}Z_t^{-1},
\end{equation}
where $Z_t$ is the renormalization constant.
We can take $g_0$ to be an arbitrary solution of the classical field
equation.  As a special case we can set $g_0=1$ (unit matrix).
We introduce the renormalization constant of the field $\pi_a$:
\begin{equation}
\pi_{a,B}= \sqrt{Z_{\pi}}\pi_a.
\end{equation}

We expand $g$ by means of $\pi_a$ as
\begin{equation}
g= g_0\Big[ 1+i\lambda T_a\pi_a-\frac{1}{2}\lambda^2(T_a\pi_a)^2
+\cdots \Big].
\end{equation}
The renomalization of the chiral model was investigated by Witten\cite{wit84}.
The correction to the term 
$(1/2t_0){\rm Tr}\partial_{\mu}g_0\partial^{\mu}g_0^{-1}$ is
\begin{equation}
\Delta\mathcal{L}= -{\rm Tr}\partial_{\mu}g_0\partial^{\mu}g_0^{-1}
\frac{C_2(G)}{8dc}\int\frac{d^dp}{(2\pi)^d}\frac{1}{p^2+m_0^2},
\end{equation}
where $C_2(G)$ is the Casimir invariant in the adjoint representation
defined by $\sum_{ab}f_{abc}f_{abd}=C_2(G)\delta_{cd}$:
\begin{equation}
C_2(G)= 2Nc~~{\rm for}~~G=SU(N).
\end{equation}
$C_2(G)$ is proportional to $(N-2)c$ for $G=O(N)$.
$m_0$ is the mass to avoid the infrared singularity.
The results are
\begin{equation}
\beta_t(\mu)\equiv \mu\frac{\partial t}{\partial\mu}= (d-2)t
-\frac{C_2(G)}{8c}t^2\frac{\Omega_d}{(2\pi)^d},
\end{equation}
for
\begin{equation}
Z_t= 1+\frac{C_2(G)}{8c}t\frac{1}{\epsilon}
\frac{\Omega_d}{(2\pi)^d}.
\end{equation}
Here $d=2-\epsilon$ and $\Omega_d=2\pi^{d/2}/\Gamma(d/2)$ is the solid angle
in $d$-dimensional space.

When $g_0$ is a constant matrix, we should expand $g$ up to the third order
of $\pi_a$ to obtain the correction to the kinetic term of $\pi_a$\cite{zin89}.
Then we obtain the beta function $\beta_t$ in a similar form\cite{zin89}.
This means that the two-point function of $\pi_a$ contains the term
$p^2/\epsilon$.

{\em Renormalization of the potential term}
Let us turn to the potential term ${\rm Tr}(g+g^{-1})$.
The renormalization is performed by expanding $g$:
\begin{equation}
{\rm Tr}g= {\rm Tr}g_0\Big[ 1+i\lambda T_a\pi_a
-\frac{1}{2}\lambda^2T_aT_b\pi_a\pi_b+\cdots \Big].
\end{equation}
The renormalization of $\alpha$ is evaluated as
\begin{eqnarray}
\alpha_0{\rm Tr}g
&\simeq& \alpha_0{\rm Tr}g_0\Big[ 1-\frac{1}{2}\lambda^2T_a^2
\langle \pi_a\pi_a\rangle +\cdots \Big]\nonumber\\
&=& \alpha_0{\rm Tr}g_0\Big[ 1-\frac{1}{2c}tT_a^2\frac{1}{\epsilon}
\frac{\Omega_d}{(2\pi)^d}+\cdots \Big],
\end{eqnarray}
up to the order of $t$.
The bare coupling constant $\alpha_0$ is related to the renormalized
$\alpha$ as follows.
\begin{equation}
\alpha_0= \alpha\mu^2 Z_{\alpha}^{-1}.
\end{equation}
Then the renormalization constant $Z_{\alpha}$ is determined as
\begin{equation}
Z_{\alpha}= 1-\frac{t}{2c}T_a^2\frac{1}{\epsilon}\frac{\Omega_d}{(2\pi)^d}
=1-\frac{t}{2c}C(N)\frac{1}{\epsilon}\frac{\Omega_d}{(2\pi)^d},
\end{equation}
where $C(N)\cdot I\equiv \sum_aT_a^2$ is the Casimir invariant in the fundamental
representation ($I$ is the unit matrix) which is given by
\begin{align}
C(N)&= c\frac{N^2-1}{N}~~{\rm for}~~G=SU(N),\\
    &= c\frac{N-1}{2}~~{\rm for}~~G=O(N),
\end{align}
The beta function for $\alpha$ is
\begin{equation}
\mu\frac{\partial\alpha}{\partial\mu}= -2\alpha
+\frac{1}{2c}t\alpha C(N)\frac{\Omega_d}{(2\pi)^d}.
\label{alphabeta}
\end{equation}
The Casimir invariant in the fundamental representation appears
in the equation of $\alpha$.


{\em Renormalization of the two-point function}
There is a contribution to the renormalization of
the coupling constant $t$ from the mass term.
We consider this up to the second order
of $\alpha$.
We set $g_0=1$ for simplicity.
We introduce the renormalization constant of
the field $\pi_a$:
\begin{equation}
\pi_{a,B}=\sqrt{Z_{\pi}}\pi_a.
\end{equation}
Using the expansion of $g$ by means of $\pi_a$, we consider the
terms that contribute to the two-point function as in the
method for the $U(1)$ sine-Gordon model\cite{ami80}.
The diagrams that contribute to the two-point function are
shown in Fig.1.

\begin{figure}
\begin{center}
\includegraphics[height=7cm,angle=90]{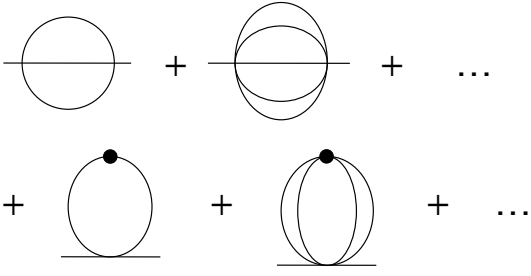}
\caption{
Contributions to the two-point function.
}
\end{center}
\label{twopoint}
\end{figure}

First, we define the Green's function:
\begin{align}
G_0({\bf x})&= \lambda^2 \langle \pi_{a}({\bf x})\pi_{a}(0)
\rangle \nonumber\\
&= \frac{1}{c} t\mu^{2-d}Z_t^{-1}Z_{\pi}^{-1}\int\frac{d^dp}{(2\pi)^d}
\frac{e^{i{\bf p}\cdot{\bf x}}}{p^2+m_0^2}\nonumber\\
&= \frac{1}{c}t\mu^{2-d}Z_t^{-1}Z_{\pi}^{-1}\frac{\Omega_d}{(2\pi)^d}K_0(m_0x),
\end{align}
where $x=|{\bf x}|$ and $K_0$ is the 0-th modified Bessel function.
$m_0$ is introduced to avoid the infrared divergence.
The first term in Fig.1 shows the contraction between
${\rm Tr}(\sum_aT_a\pi_ai(x))^4$ and ${\rm Tr}(\sum_bT_b\pi_b(y))^4$. 
This term gives $G_0({\bf x})^3$ and its coefficient is given by
$\sum_{bcd}{\rm Tr}T_aT_bT_cT_d\times {\rm Tr}(T_aT_bT_cT_d)$ where
${\rm Tr}(T_aT_bT_cT_d)$ includes the trace of permutations of
$T_a$, $T_b$, $T_c$ and $T_d$.
We estimate this by using the relation for $SU(N)$ given by
\begin{equation}
T_aT_b= \frac{1}{2N}\delta_{ab}\cdot I+\frac{1}{2}
\sum_{c=1}^{N^2-1}(if_{abc}+d_{abc})T_c,
\end{equation}
for $c=1/2$.
$\{d_{abc}\}$ are structure constants and are totally symmetric
with respect to indices $a$, $b$ and $c$.
Because we include the permutations of $T_aT_bT_cT_d$, we drop
$f_{abc}$ term in this formula.  We find that the coefficient of
the first term in Fig.1 is of order $N^2$ for large $N$ by
using the relation
\begin{equation}
\sum_{ab}d_{abc}d_{abd}= (N^2-4)/N\cdot \delta_{cd}.
\end{equation}

Since ${\rm sinh}I-I=I^3/3!+\cdots$,
the diagrams in Fig.1 are summed up to give\cite{ami80}
\begin{equation}
\Sigma 
= \int d^d x\Big[ e^{i{\bf p}\cdot{\bf x}}\left({\rm sinh}I-I\right)
-\left( {\rm cosh}I-1\right)\Big],
\end{equation}
where $I=C(N)G_0(x)$.  
We use ${\rm sinh}I-I\approx (1/2)e^I$ and 
${\rm cosh}I \approx (1/2)e^I$ for large $I$.
We included $C(N)$ in $I$.
We define 
$A_0\equiv\sum_{bcd}{\rm Tr}T_aT_bT_cT_d\times{\rm Tr}(T_aT_bT_cT_d)/
C(N)^3$ which is of order $1/N$ for large $N$.
Then, the correction to the bare two-point function for $\pi$ fields is given as
\begin{align}
\Gamma_B^{(2)c}(p)&= -\frac{1}{2}\left( \frac{\alpha}{t}\mu^d 
Z_{\alpha}^{-1}Z_t \right)^2 A_0\frac{t\mu^{2-d}}{Z_tZ_{\pi}} 
\int d^d x\left(e^{i{\bf p}\cdot{\bf x}}-1\right) \nonumber\\
&\times e^{C(N)G_0(x)}.
\end{align}
We use 
$e^{i{\bf p}\cdot{\bf x}}=1+i{\bf p}\cdot{\bf x}-(1/2)({\bf p}\cdot{\bf x})^2+\cdots$
and keep the ${\bf p}^2$ term.  
We use the asymptotic formula $K_0(x)\sim -\gamma-\log(x/2)$ for small $x$.
Because $t$ has a fixed point $t_c$ from a zero of eq.(\ref{alphabeta}),
we denote by $v$ the deviation of $t$ from the fixed point:
\begin{equation}
\frac{1}{8\pi c}C(N)t=1+v,
\end{equation}
where we set $d=2$ for the volume element $\Omega_d/(2\pi)^d$.
The bare two-point function reads
\begin{align}
\Gamma_B^{(2)c}(p)&= \frac{1}{8}A_0\frac{\alpha^2}{t}\mu^{d+2}p^2
(c_0m_0^2)^{-2-2v}\Omega_d \frac{Z_t}{Z_{\pi}}\nonumber\\
&\times \int_0^{\infty}dx x^{d+1} 
\frac{1}{(x^2+a^2)^{2+2v}}\nonumber\\
&= \frac{1}{8}A_0p^2 \frac{\alpha^2}{t}\mu^{d+2}(c_0m_0^2)^{-2}
\frac{Z_t}{Z_{\pi}}\Omega_d
\frac{1}{\epsilon}\nonumber\\
&\simeq p^2 \frac{1}{32c}A_0C(N)\alpha^2\mu^{d+2}(c_0m_0^2)^{-2}
\frac{Z_t}{Z_{\pi}}\frac{1}{\epsilon}+O(v), \nonumber\\
\end{align} 
where
$c_0$ is a constant and $a=1/\mu$ is a small cutoff to keep away from a
infrared singularity.
The divergence of $\alpha$ was absorbed by $Z_{\alpha}$.
The renormalized two-point function is given by
\begin{eqnarray}
\Gamma_R^{(2)}(p)&=& Z_{\pi}\Gamma_B^{(2)}(p) \nonumber\\
&=& \frac{cZ_tZ_{\pi}}{t\mu^{2-d}}\Big[ p^2+p^2\frac{1}{32c}A_0C(N)\alpha^2\mu^{d+2} \nonumber\\
&& \times (c_0m_0^2)^{-2}\frac{1}{\epsilon} \Big].
\end{eqnarray}
The divergence is removed by $Z_t$.
For $g_0=1$ (constant matrix),
$\Gamma_R^{(2)}(p)$ contains the term $p^2/\epsilon$ coming
from the kinetic part of the Lagrangian.
The $\alpha^2$-term of $Z_t$ is determined as
\begin{equation}
Z_t= 1-\frac{1}{32c}A_0C(N)\alpha^2\mu^{d+2}(c_0m_0^2)^{-2}
\frac{1}{\epsilon}.
\end{equation}
From the equation $\mu\partial t_0/\partial\mu=0$ for $t_0=Z_t^{-1}t\mu^{2-d}$,
we obtain the contribution of order $\alpha^2$ as
\begin{align}
\mu\frac{\partial t}{\partial\mu}&=(d-2)t-\frac{1}{32c}A_0C(N)(c_0m_0^2)^{-2}
t\frac{1}{\epsilon}\nonumber\\
&\times \left( 2\alpha\mu^{d+2}\mu
\frac{\partial\alpha}{\partial\mu}+(d+2)\alpha^2\mu^{d+2}\right).
\end{align}
With use of eq.(\ref{alphabeta}) this results in
\begin{align}
\mu\frac{\partial t}{\partial\mu}&= (d-2)t+\frac{1}{32c}A_0C(N)\mu^{d+2}
(c_0m_0^2)^{-2}t\alpha^2 \frac{\Omega_d}{(2\pi)^d}\nonumber\\
&+O(\alpha^2 t^2).
\end{align}
We perform the finite renormalization for $\alpha$ in the 
manner\cite{ami80}
$\alpha\rightarrow \alpha c_0m_0^2 a^2=\alpha c_0m_0^2\mu^{-2}$
because the finite part of $G_0(x\rightarrow 0)$ is 
$G_0(x\rightarrow 0)=-(1/2\pi)\ln(m_0/\mu)$.
We add the result for the chiral Lagrangian to obtain scaling
equations:

\begin{align}
\mu\frac{\partial t}{\partial\mu}&= (d-2)t-\frac{C_2(G)}{8c}t^2
+\frac{1}{32c}A_0C(N)t\alpha^2,\\
\mu\frac{\partial \alpha}{\partial\mu}&= -\alpha\left( 2-C(N)\frac{1}{2c}
t\right), 
\end{align}
wherae we include $\Omega_d/(2\pi)^d$ into the definition of $t$ for simplicity.


{\bf Renormalization flow and asymptotic freedom}

Let us consider the renormalization group flow in two dimensions.
Since $A_0\sim O(1/N)$,
we set $A_0=\tilde{A_0}/N$ and we define 
$\tilde{\alpha}=\sqrt{\tilde{A_0}/32}\alpha$.
Then the equations read
\begin{align}
\mu\frac{\partial t}{\partial\mu}&= (d-2)t-\frac{C_2(G)}{8c}t^2
+\frac{1}{cN}C(N)t\tilde{\alpha}^2,\\
\mu\frac{\partial\tilde{\alpha}}{\partial\mu}&= 
-\tilde{\alpha}\left( 2-\frac{C(N)}{2c}t\right).
\end{align}
In the following, we write $\tilde{\alpha}$ as $\alpha$ for simplicity.

The beta functions for $SU(N)$ with $N\ge 2$ are
\begin{align}
\mu\frac{\partial t}{\partial\mu}&= (d-2)t-\frac{N}{4}t^2
+\frac{N^2-1}{N^2}t\alpha^2,\\
\mu\frac{\partial\alpha}{\partial\mu}&= 
-\alpha\left( 2-\frac{N^2-1}{2N}t\right).
\end{align}
This set of equations has several phases depending on the initial
values of $t$ and $\alpha$.

{\em Bifurcation point}
A set of beta functions has zero at $(t,\alpha)= (t_c,\alpha_c)$
in two dimensions $d=2$ where
\begin{equation}
t_c = \frac{4c}{C(N)},~~~
\alpha_c =\sqrt{\frac{cNC_2(G)}{2}}\frac{1}{C(N)}.
\end{equation}
This point, however, divides the parameter space of $(t,\alpha)$
into two regions.  One is the strong coupling region and the other
is the weak coupling region.


\begin{figure}
\begin{center}
\includegraphics[height=6cm]{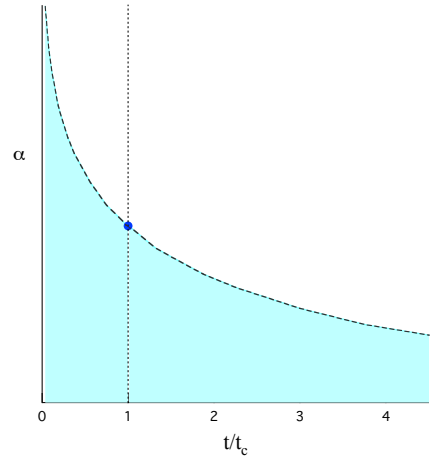}
\caption{
The region (shaded region except for the boundary) where 
the flow is renormalized into 
$\mu\partial t/\partial\mu<0$ and
$\mu\partial\alpha/\partial\mu<0$, so that $t\rightarrow 0$ and
$\alpha\rightarrow 0$ as $\mu$ increases.
Outside of this region, the flow goes into the strong-coupling
region given by $t\rightarrow\infty$ and $\alpha\rightarrow\infty$
as $\mu$ increases.
The points on the boundary go to the critical point 
$(t_c,\alpha_c)$
along the boundary following the renormalization flow. 
}
\end{center}
\label{t-alpha}
\end{figure}

{\em Asymptotic freedom}
When $\mu\partial t/\partial\mu<0$ and $\mu\partial\alpha/\partial\mu<0$,
we have an asymptotically free theory.  
If the parameters $(t,\alpha)$ goes into the
region,
\begin{equation}
t<t_c,~~~\frac{C(N)}{N}\alpha^2< \frac{C_2(G)}{8}t,
\label{ineq}
\end{equation}
$t$ approaches 0 as $\mu$ increases.
When the flow goes into this region, the model shows an asymptotic
freedom.
A parameter $(t,\alpha)$ near this region will also go into 
an asymptotically free region.
We show the region of the initial values of $(t,\alpha)$ in Fig.2 
where
the flows are renormalized into the region of asymptotic freedom with
$t\rightarrow 0$ and $\alpha\rightarrow 0$ as $\mu\rightarrow\infty$.
Outside of the shaded region, the flow goes to the strong-coupling
region, that is, $t\rightarrow\infty$ and $\alpha\rightarrow\infty$
as $\mu$ increases.
When the initial parameters $(t,\alpha)$ are on the boundary, 
they are renormalized into $(t_c,\alpha_c)$.

{\em $SU(2)$ sine-Gordon model}
We solve a set of scaling equations numerically for the $SU(2)$ case
($N=2$).  In this case we have $(t_c,\alpha_c)=(8/3,16/3)$.
We show the renormalization group flow in Fig.3.
Obviously we have the weak coupling and strong coupling regimes.
The former is the asymptotically free region where $t$ and
$\alpha$ are renormalized to zero. 
The divide between two regions is shown in Fig.4 by the dashed
line.  All the renormalization flows approach a line as
$\mu\rightarrow\infty$, which is shown by the solid line in
Fig.4.

In the strong coupling region where $t$ and $\alpha$ are large,
the Lagrangian may be approximated by the mass term:
$\mathcal{L}\simeq \alpha/(2t){\rm Tr}(g+g^{-1})$.
We have mass gap in this region.
Instead, in the weak coupling region where both $t$ and $\alpha$
approach zero, the Lagrangian is effectively given by the
kinetic term because $\alpha$ is decreased faster than $t$.

\begin{figure}
\begin{center}
\includegraphics[height=6.5cm]{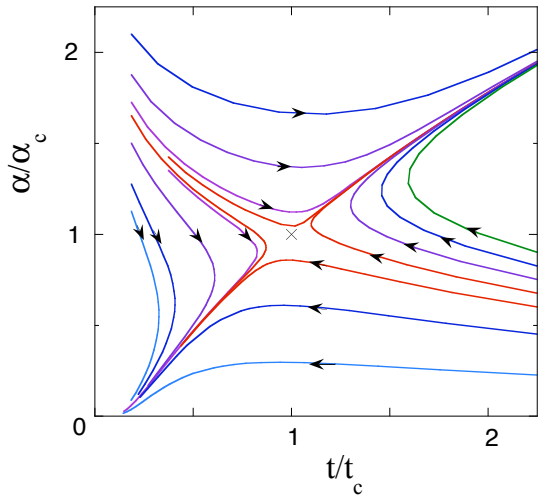}
\caption{
Renormalization flow as $\mu$ increases.
The cross indicates the fixed point $(t_c,\alpha_c)$.
}
\end{center}
\label{flow}
\end{figure}

\begin{figure}
\begin{center}
\includegraphics[height=6.5cm]{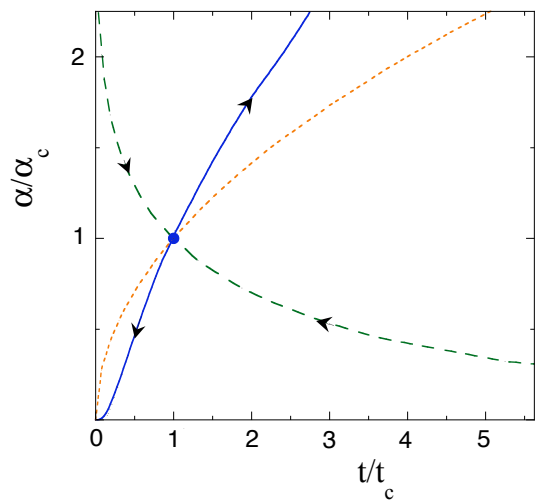}
\caption{
Renormalization flow approaches the solid line as $\mu$ 
increases.  The dotted line indicates $\alpha=\sqrt{2t/3}$
below which we have $\mu\partial t/\partial\mu<0$.
The dashed line represents a ridge that divides the $t-\alpha$
plane into two regions.
}
\end{center}
\label{asympline}
\end{figure}

{\bf A matrix model and large-$N$ limit}
In the limit of large $t$ and $\alpha/t$,
the $SU(N)$ sine-Gordon model is reduced to a unitary matrix model.
A third-order phase transition has been predicted for this 
model\cite{gro80, bre80}.
Gross and Witten considered the partition function of the form
\begin{equation}
Z= \int dU\exp\left(\beta N{\rm Tr}(U+U^{\dag})\right),
\label{matrix}
\end{equation}
where $U$ is, for example, an $N\times N$ special unitary matrix, 
$UU^{\dag}=1$.
There is a phase transition at $\beta=1/2$ in the large $N$ limit.
This is a transition between the weak coupling regime $\beta >1/2$
and the strong coupling regime $\beta < 1/2$. 
It has been shown that this is a third-order transition because
the third derivative of the free energy, $-\ln Z$, is discontinuous
at $\beta=1/2$.
 
To consider the relation with the unitary matrix model, we replace
$\alpha$ to $N\alpha$ as in eq.(\ref{matrix}).  
When $N$ is large, the $SU(N)$ beta functions are reduced to
\begin{equation}
\mu\frac{\partial t}{\partial\mu}=
N^2t\alpha^2,~~
\mu\frac{\partial\alpha}{\partial\mu}=
-\alpha\left( 2-\frac{N}{2}t\right).
\end{equation}
The zero of $\mu\partial\alpha/\partial t=0$ gives
the fixed point
$t_c\simeq 4/N$.
Near this point, $t$ is parametrized as $t=(4/N)(1+v)$.  Then the equations
read
\begin{equation}
\mu\frac{\partial v}{\partial\mu}= \frac{N^2}{2}\alpha^2+O(v\alpha^2),~~
\mu\frac{\partial\alpha}{\partial\mu}=2\alpha v,
\end{equation}
where we neglected the term $v\alpha^2$.
When $v\alpha^2$ is small,
a set of equations was reduced to that of the Kosterlitz-Thouless
transition\cite{kos73,zin89},
namely, that of the Abelian ($U(1)$) sine-Gordon model.
We define $x=2v$ and $y=N\alpha$ to obtain\cite{kos73}
$\mu\partial x/\partial \mu= y^2$,
$\mu\partial y/\partial\mu= xy$, i.e.,
the renormalization group flow, which is identical to that of the
Kosterlitz-Thouless transition.

We define $s\equiv t/t_c\simeq (N/4)t$.  
The Lagrangian is written as
\begin{equation}
\mathcal{L}= \frac{N}{8s}{\rm Tr}\partial_{\mu}g\partial^{\mu}g^{-1}
+\frac{N\alpha}{8s}{\rm Tr}(g+g^{-1}).
\end{equation}
In the strong coupling region, $s\gg 1$ and $\alpha\gg 1$, 
$\mathcal{L}$ is approximated by a matrix model
$\mathcal{L}_m = N\alpha/(8s){\rm Tr}(g+g^{-1})$.
$\alpha/(8s)$ corresponds to $\beta$ and $\alpha/s=4$ is a divide
of strong and weak coupling regimes.
The renormalization flow in the large $N$ limit is shown in Fig.5,
where the flow goes into the strong coupling region as $\mu$ 
increases.

\begin{figure}
\begin{center}
\includegraphics[height=6.0cm]{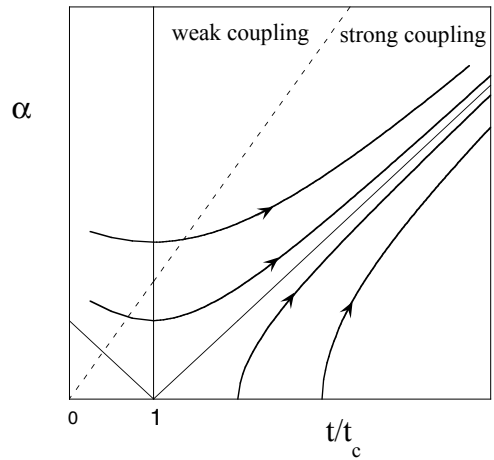}
\caption{
The renormalization group flow for large $N$.
The dashed line indicates $\alpha/s=4$ where $s=t/t_c=1+v$.
}
\end{center}
\label{rg-largen}
\end{figure}

{\bf Discussion}
Our model is a generalization of the Abelian $U(1)$
sine-Gordon model to the model with gauge group $G$, and is also regarded 
as a generalization of the chiral model with a mass term. 
This model is also regarded as a nonabelian generalization of the Josephson model
in superconductors.
The nonabelian sine-Gordon model is a multi-component field theory 
and each independent component in $g$ represents a mode in a periodic potential.

When we take $g=e^{-i\varphi\sigma_3/2}e^{i\theta\sigma_2/2}e^{i\varphi\sigma_3/2}$,
the Lagrangian is
$\mathcal{L}= (1/4t)\Big[ (\partial_{\mu}\theta)^2
+4\sin^2\theta(\partial_{\mu}\varphi)^2 \Big]+(2\alpha/t)\cos(\theta/2)$.
The kinetic term is that for the nonlinear sigma model with
${\bf n}=(\sin\theta\cos(2\varphi),\sin\theta\sin(2\varphi),\cos\theta)$.
This model has one massive and one massless mode, if we replace $\sin^2\theta$
by its expectation value using a mean-field like treatment.
We can generalize the model to have multiple massive modes and multiple
massless Nambu-Goldstone modes.
We can add an external field or a higher oder potential term such as 
$\alpha_2{\rm Tr}g^2$ to the Lagrangian:
\begin{equation}
\mathcal{L}= \frac{1}{4t}\Big[ (\partial_{\mu}\theta)^2
+4\sin^2\theta(\partial_{\mu}\varphi)^2 \Big]+\frac{2\alpha}{t}\cos\left(
\frac{\theta}{2}\right)+\frac{2\alpha_2}{t}\cos\theta.
\end{equation}
The potential term may have a non-trivial ground state with finite $\theta$
depending on the sign of $\alpha$ and $\alpha_2$.
In the state with a finite stationary value of $\theta$, the ground states
are degenerate, leading to the breaking of time-reversal symmetry 
(or {\em CP} invariance)\cite{yan14}.
There is a transition as $\alpha_2$ is varied.
We expect that this kind of transition may occur in an
unconventional superconductor\cite{tan14}.  

For $g=e^{i\phi\sigma_2/2}e^{i\varphi\sigma_3/2}\in SU(2)$, the Lagrangian reads
\begin{align}
\mathcal{L}&= \frac{1}{4t}\left( (\partial_{\mu}\phi)^2
+(\partial_{\mu}\varphi)^2 \right)+\frac{2\alpha}{t}\cos\left(\frac{\phi}{2}\right)
\cos\left(\frac{\varphi}{2}\right)\nonumber\\
&+ \frac{\alpha_2}{t}(\cos\phi\cos\varphi+\cos\phi+\cos\varphi-1),
\end{align}
where we added an external field Tr$(g^2+g^{-2})$.
This is a model of coupled scalar fields, where both fields are massive.
There would also be a transition when $\alpha$ and $\alpha_2$ terms are
frustrated.

{\bf Summary}
We investigated the scaling property of the chiral
sine-Gordon model with $G$-valued fields for $G=SU(N)$.
We derived a set of renormalization group equations for 
this model, where the coefficients of the beta functions are
determined by Casimir invariants of $G$.
There are two regions in the parameter space of $t$ and $\alpha$: 
one is the ultraviolet strong-coupling region
and the other is the asymptotically free region.
The beta functions have zero at $(t,\alpha)=(t_c,\alpha_c)$ where
the model has scale invariance.
This point divides the parameter space into
two regions.

We considered the large-$N$ model.
The beta functions in this model are simplified and reduced to 
those for the 
Kosterlitz-Thouless transition, that is, the $U(1)$ sine-Gordon model
 near the critical point.
In the strong coupling limit, the $SU(N)$ model is reduced to a matrix model
with $U(N)$-value fields.  There may be a third-order phase
transition at $\alpha/s=4$ in the large-$N$ limit.


The author expresses his sincere thanks to Prof. S. Hikami 
for stimulating  discussions.
This work was supported in part by Grant-in-Aid from the
Ministry of Education and Science of Japan (Grant No. 22540381).

\end{document}